# The Wave Function and Quantum Reality

Shan Gao

*Unit for History and Philosophy of Science & Centre for Time, SOPHI*
*University of Sydney, Sydney, NSW 2006, Australia*

**Abstract.** We investigate the meaning of the wave function by analyzing the mass and charge density distribution of a quantum system. According to protective measurement, a charged quantum system has mass and charge density proportional to the modulus square of its wave function. It is shown that the mass and charge density is not real but effective, and it is formed by the ergodic motion of a localized particle with the total mass and charge of the system. Moreover, it is argued that the ergodic motion is not continuous but discontinuous and random. This result suggests a new interpretation of the wave function, according to which the wave function is a description of random discontinuous motion of particles, and the modulus square of the wave function gives the probability density of the particles *being* in certain locations. It is shown that the suggested interpretation of the wave function disfavors the de Broglie-Bohm theory and the many-worlds interpretation but favors the dynamical collapse theories, and the random discontinuous motion of particles may provide an appropriate random source to collapse the wave function.

## 1. Introduction

The wave function is the most fundamental concept of quantum mechanics. According to the standard probability interpretation, the wave function is a probability amplitude, and its modulus square gives the probability density of particles being found in certain locations. However, this interpretation is not wholly satisfactory because of resorting to the vague concept - measurement (Bell 1990). On the other hand, the meaning of the wave function is also in dispute in the alternative interpretations of quantum mechanics such as the de Broglie-Bohm theory and the many-worlds interpretation (de Broglie 1928; Bohm 1952; Everett 1957). In view of this unsatisfactory situation, it seems that we need a new starting point to solve this fundamental interpretive problem of quantum mechanics.

In this paper[1], we will analyze the possible implications of protective measurement for the meaning of the wave function. Like conventional impulse measurement, protective measurement also uses a standard measuring procedure, but with a weak and long duration coupling interaction. Besides, it adds an appropriate procedure to protect the measured wave function from being changed (in some situations the protection is provided by the measured system itself). These differences permit protective measurement to be able to gain more information about the measured quantum system and its wave function, and thus it might unveil more physical content of the wave function.

## 2. Protective measurement and charge density

According to protective measurement, the mass and charge density of a single quantum system can be measured as expectation values of certain variables. It turns out that the mass and the charge of a charged quantum system are distributed throughout space with density proportional to the modulus square of its wave function. (Aharonov and Vaidman 1993; Aharonov, Anandan and Vaidman 1993; Aharonov, Anandan and Vaidman 1996).

---
[1] An enlarged version of this paper is available at PhilSci-Archive (Gao 2011).

Consider a quantum system in a discrete nondegenerate energy eigenstate $\psi(x)$. A protective measurement of an observable $A_n$, which is a normalized projection operator on small regions $V_n$ having volume $v_n$, will yield the following result:

$$\langle A_n \rangle = \frac{1}{v_n} \int_{v_n} |\psi(x)|^2 dv = |\psi_n|^2 \qquad (1)$$

It is the average of the density $|\psi(x)|^2$ over the small region $V_n$. When $v_n \to 0$ and after performing measurements in sufficiently many regions $V_n$ we can find the whole density distribution $|\psi(x)|^2$. For a charged system with charge $Q$, the density $|\psi(x)|^2$ times the charge yields the charge density $Q|\psi(x)|^2$. In particular, an appropriate adiabatic measurement of the Gauss flux out of a certain region will yield the value of the total charge inside this region, namely the integral of the charge density $Q|\psi(x)|^2$ over this region. Similarly, we can measure the mass density of the system in principle by an appropriate adiabatic measurement of the flux of its gravitational field in principle.

In a word, protective measurement shows that the mass and charge of a single quantum system, which is described by the wave function $\psi(x)$, is distributed throughout space with mass density $m|\psi(x)|^2$ and charge density $Q|\psi(x)|^2$, respectively[2].

## 3. The physical origin of charge density

The key to unveil the meaning of the wave function is to find the physical origin of the mass and charge density. If the mass and charge density of a charged quantum system is real, that is, if the densities at different locations exist at the same time, then there will exist gravitational and electrostatic self-interactions of the density, and the Schrödinger equation for a free quantum system with mass $m$ and charge $Q$ will be

$$i\hbar \frac{\partial \psi(x,t)}{\partial t} = -\frac{\hbar^2}{2m} \nabla^2 \psi(x,t) + (kQ^2 - Gm^2) \int \frac{|\psi(x',t)|^2}{|x-x'|} d^3x' \psi(x,t) \qquad (2)$$

where $k$ is the Coulomb constant, and $G$ is Newton's gravitational constant.

It has been shown that the measure of the potential strength of a gravitational self-interaction is $\varepsilon^2 = (4Gm^2/\hbar c)^2$ for a free system with mass $m$ (Salzman 2005). This quantity represents the strength of the influence of self-interaction on the normal evolution of the wave function; when $\varepsilon^2 \approx 1$ the influence will be significant. Similarly, for a free charged system with charge $Q$, the measure of the potential strength of the electrostatic self-interaction is $\varepsilon^2 = (4kQ^2/\hbar c)^2$. For example, the potential strength of the electrostatic self-interaction is $\varepsilon^2 = (4ke^2/\hbar c)^2 \approx 1 \times 10^{-3}$ for a free electron. This indicates that the electrostatic self-interaction will have a remarkable influence on the evolution of its wave function[3]. If such an interaction indeed exists, it should have been detected by precise interference experiments on electrons. On the other hand, the superposition principle of quantum mechanics, which denies the existence of the observable electrostatic self-interaction, has been verified for microscopic particles with astonishing precision. As another example, consider the electron in the hydrogen atom. Since the potential of the electrostatic self-interaction is of the same order as the Coulomb potential produced by the nucleus, the energy levels of hydrogen atoms will be remarkably different from those predicted by quantum mechanics and confirmed by experiments.

---

[2] For a more detailed introduction of this result see Gao (2011).
[3] By contrast, the potential strength of the gravitational self-interaction for a free electron is $\varepsilon^2 = (4Gm_e^2/\hbar c)^2 \approx 4 \times 10^{-89}$.

Therefore, although the gravitational self-interaction is too weak to be detected presently, the existence of the electrostatic self-interaction for a charged quantum system such as an electron already contradicts experimental observations. This means that the mass and charge density of a quantum system cannot be real but be effective; at every instant there is only a localized particle with the total mass and charge of the system, and during a time interval the time average of the ergodic motion of the particle forms the effective mass and charge density. There exist no gravitational and electrostatic self-interactions of the density in this case.

It can be further argued that the classical ergodic models, which assume continuous motion of particles, are inconsistent with quantum mechanics (Gao 2011). For example, consider an electron in a superposition of two energy eigenstates in two boxes. Even if one assumes that the electron can move with infinite velocity (e.g. at the nodes), due to the restriction of box walls it cannot continuously move from one box to another and generate the required charge density proportional to the modulus square of the superposition state. Therefore, the ergodic motion of particles cannot be continuous but be discontinuous. If the motion of a particle is discontinuous, then the particle can readily move throughout all regions where the wave function is nonzero during an arbitrarily short time interval at a given instant. Moreover, if the probability density of the particle appearing in each position is proportional to the modulus square of its wave function there at every instant, the discontinuous motion can also generate the right effective mass and charge density.

To sum up, the mass and charge density of a quantum system, which is measurable by protective measurement, is not real but effective. Moreover, the effective mass and charge density is formed by the discontinuous motion of a localized particle, and the probability density of the particle appearing in each position is proportional to the modulus square of its wave function there. As a result, the wave function of the system can be regarded as a description of the random discontinuous motion of the localized particle.

## 4. The wave function as a description of random discontinuous motion of particles

In classical mechanics, we have a clear physical picture of motion. It is well understood that the trajectory function $x(t)$ in classical mechanics describes the continuous motion of a particle. In quantum mechanics, the trajectory function $x(t)$ is replaced by a wave function $\psi(x,t)$. Since quantum mechanics is a more fundamental theory of the physical world, of which classical mechanics is only an approximation, it seems natural that the wave function should describe some sort of more fundamental motion of particles, of which continuous motion is a only an approximation in the classical domain. The above analysis provides a strong support for this conjecture, and it suggests that what the wave function describes is probably the more fundamental motion of particles, which is essentially discontinuous and random.

The physical definition of random discontinuous motion of a particle is as follows. The position of the particle at each instant is only determined by a certain instantaneous condition at the instant in a probabilistic way, and this probabilistic instantaneous condition gives the probability density of the particle appearing in every position in space. As a result, the trajectory of the particle is essentially discontinuous, i.e., that the trajectory function $x(t)$ of the particle is not continuous at every instant. The strict mathematical description of random discontinuous motion can be obtained by using the measure theory in mathematics (see, e.g. Cohn 1993). It has been shown that the position density $\rho(x,t)$ and the position flux density $j(x,t)$ provide a complete description for the random discontinuous motion of a single particle (Gao 2011). By assuming that the nonrelativistic equation of motion is the Schrödinger equation, the wave function $\psi(x,t)$ can be uniquely expressed by $\rho(x,t)$ and $j(x,t)$, and it also provides a complete description of the state of random discontinuous motion of a single particle. For the motion of many particles, the joint position density and joint position flux density are defined in the 3N-dimensional configuration space, and thus the many-particle wave function, which is composed of these two quantities, is also defined in the 3N-dimensional configuration space.

Interestingly, we can reverse the above logic in some sense, namely by assuming the wave function is a complete description for the motion of particles, we can also reach the random discontinuous motion of particles, independent of our previous analysis. If the wave function $\psi(x,t)$ is a description of the state of motion for a single particle, then the quantity $|\psi(x,t)|^2 dx$ not only gives the probability of the particle being found in an infinitesimal space interval $dx$ near position $x$ at instant $t$ (as in standard quantum mechanics), but also gives the objective probability of the particle being there. This accords with the common-sense assumption that the probability distribution of the measurement results of a property is the same as the objective distribution of the property in the measured state. Then at instant $t$ the particle may appear in any location where the probability density $|\psi(x,t)|^2$ is nonzero, and during an infinitesimal time interval near instant $t$, the particle will move throughout the whole space where the wave function $\psi(x,t)$ spreads, though it is still in one position at each instant. Moreover, its position density is equal to the probability density $|\psi(x,t)|^2$. Obviously this kind of motion is essentially random and discontinuous.

One important point needs to be stressed here. Since the wave function in quantum mechanics is defined at an instant, not during an infinitesimal time interval, it should be regarded not as a description of the state of random discontinuous motion of particles, but as a description of the instantaneous condition or instantaneous intrinsic property of the particles that determines their random discontinuous motion at a deeper level[4]. In particular, the modulus square of the wave function determines the probability density of the particles appearing in every position in space at a given instant. This intrinsic property may be called indeterministic disposition or propensity[5]. By contrast, the position density $\rho(x,t)$ and the position flux density $j(x,t)$, which are defined during an infinitesimal time interval, are only a description of the state of the resulting random discontinuous motion of particles, and they are determined by the wave function. In this sense, we may say that the motion of particles is "guided" by the wave function in a probabilistic way.

The suggested interpretation of the wave function in terms of random discontinuous motion of particles might be taken as a natural realistic extension of the orthodox view. The naturalness of the extension lies in that it still makes particles ontological and the wave function epistemological[6]. That the extension is realistic is obvious. According to Born's probability interpretation, the modulus square of the wave function of a particle gives the probability density of the particle *being found* in certain positions, while according to the suggested interpretation, the modulus square of the wave function also gives the objective probability density of the particle *being* there. Certainly, the transition process from "being" to "being found", which is closely related to the quantum measurement problem, needs to be further explained. We will discuss this important issue in the next section.

## 5. Implications for the solution to the measurement problem

In standard quantum mechanics, it is postulated that when a wave function is measured by a macroscopic device, it will no longer follow the linear Schrödinger equation, but instantaneously collapse to one of the wave functions that correspond to definite measurement results. However, this collapse postulate is only a makeshift, and the theory does not tell us why and how the definite measurement result appears (Bell 1990). There are in general two ways to solve the measurement problem. The first one is to integrate the collapse evolution with the normal Schrödinger evolution into a unified dynamics, e.g. in the dynamical collapse theories (Ghirardi 2008). The second way is to reject the postulate and assume that the Schrödinger equation completely describes the evolution of the wave function. There are two main alternative theories along this avoiding-collapse direction. The first one is the de Broglie-Bohm theory (de Broglie 1928; Bohm 1952), which takes the wave function as an incomplete description and adds some hidden variables to explain the definite measurement results. The second one is the many-worlds interpretation (Everett 1957), which assumes the existence of many worlds to explain our definite experience in one of these worlds and still regards the wave function as a complete description of the whole worlds. In this section, we will analyze the possible implications of our suggested interpretation of the wave function for these solutions to the measurement problem.

---

[4] From a logical point of view, for the random discontinuous motion of particles, the particles should also have an intrinsic property that determines their discontinuous motion in a probabilistic way, otherwise they would not "know" how frequently they should appear in every position in space. See also the definition of random discontinuous motion given in the last section.
[5] It is worth noting that this kind of propensity relates to the objective motion of particles, not to the measurement on the particles (cf. Suárez 2004).
[6] By contrast, the de Broglie-Bohm theory and the many-worlds interpretation both attach reality to the wave function itself (Bohm 1952; Everett 1957).

At first sight, the above three theories seem apparently inconsistent with the suggested interpretation of the wave function. They all attach reality to the wave function, e.g. taking the wave function as a real physical entity on configuration space or assuming the wave function has a field-like spatiotemporal manifestation in the real three-dimensional space (see, e.g. Ghirardi 1997, 2008; Wallace and Timpson 2009). But according to our suggested interpretation, the wave function is not a field-like physical entity on configuration space; rather, it is a description of the random discontinuous motion of particles in three-dimensional space (and at a deeper level a description of the instantaneous intrinsic property of the particles that determines their random discontinuous motion). Anyway, in spite of the various views on the wave function in these theories, they never interpret the wave function as a description of the motion of particles in three-dimensional space. However, on the one hand, the interpretation of the wave function in these theories is still an unsettled issue, and on the other hand, these theories may be not influenced by the interpretation in a significant way. Therefore, they may be consistent with the suggested interpretation of the wave function after certain revision.

*5.1 The de Broglie-Bohm theory*

Let's first analyze the de Broglie-Bohm theory (de Broglie 1928; Bohm 1952). According to the theory, a complete realistic description of a quantum system is provided by the configuration defined by the position of its particle together with its wave function. The wave function follows the linear Schrödinger equation and never collapses. The particle, often called Bohmian particle, is guided by the wave function via the guiding equation to undergo continuous motion. The result of a measurement is indicated by the position of the Bohmian particle describing the pointer of the measuring device, and thus it is always definite. Moreover, it can be shown that the de Broglie-Bohm theory gives the same predictions of measurement results as standard quantum mechanics by means of a quantum equilibrium hypothesis (so long as the latter gives unambiguous predictions). In this way, it seems that the de Broglie-Bohm theory can succeed in avoiding the collapse of the wave function.

However, although the de Broglie-Bohm theory is mathematically equivalent to standard quantum mechanics, there is no clear consensus with regard to its physical interpretation. To begin with, the interpretation of the wave function in the theory is still in dispute. For example, the wave function has been regarded as a field similar to electromagnetic field (Bohm 1952), an active information field (Bohm and Hiley 1993), a field carrying energy and momentum (Holland 1993), a causal agent more abstract than ordinary fields (Valentini 1997), and a component of physical law (Dürr, Goldstein and Zanghì 1997; Goldstein and Teufel 2001) etc. Notwithstanding the differences between these interpretations, they are inconsistent with the picture of random discontinuous motion of particles. To say the least, they can hardly explain the existence of mass and charge density for a charged quantum system, which is measurable by protective measurement. Our previous analysis suggests that the mass and charge density of a quantum system, which is proportional to the modulus square of its wave function, is effective and formed by the ergodic motion of a localized particle with the total mass and charge of the system, and thus the wave function is a description of the ergodic motion of particles.

Next, let's analyze the hypothetical Bohmian particles in the de Broglie-Bohm theory. It has been argued that the mass and charge of a quantum system should be possessed by its wave function, not by its Bohmian particle (see, e.g. Brown, Dewdney and Horton 1995). It is even claimed that a Bohmian particle has no properties other than its position (Hanson and Thoma 2011). As our previous analysis suggests, protective measurement may provide a more convincing argument for the "bareness" of the Bohmian particles. The existence of mass and charge density for a charged quantum system, which is proportional to the modulus square of its wave function and measurable by protective measurement, implies that mass and charge are attributes of the wavefunction and not of the hypothetical Bohmian particle. When the wave function is further interpreted as a description of the random discontinuous motion of particles as we have suggested, it becomes more obvious that the mass and charge (and other properties) of a quantum system belong to these particles, not to the added Bohmian particles.

The "bareness" of the Bohmian particles is at least a worrisome issue. According to the common-sense view, a real particle should have its intrinsic properties such as mass and charge etc, and its total energy cannot be zero either. If a particle has no properties other than its position, then in what sense it can be regarded as physically real? It seems that a bare Bohmian particle has no difference with a mathematical point. Furthermore, if the Bohmian particles are deprived of all intrinsic properties, then how can they be guided by the wave function? and how can the wave function "know" its existence and guide its motion? This also reminds us another debatable aspect of the de Broglie-Bohm theory, the interaction between the wave function and the Bohmian particles. In the final analysis, the influence of the wave function on the Bohmian particles is in want of a physical explanation.

To sum up, when taking into account of the implications of protective measurement and our suggested interpretation of the wave function based on them, the de Broglie-Bohm theory seems to be not a satisfactory solution to the measurement problem[7]. Although the theory can be mathematically equivalent to standard quantum mechanics, it seems lack of a reasonable physical interpretation. The added hidden variables, which are used to explain the emergence of definite measurement results, can only be carried by bare particles without any intrinsic properties of the involved quantum system such as mass and charge. Moreover, the theory can hardly explain why the evolution of the hidden variables is guided by the wave function in the way it requires. In particular, when the wave function is interpreted as a description of the random discontinuous motion of particles (and at a deeper level a description of the intrinsic property of the particles that determines their discontinuous motion in a probabilistic way), it seems impossible that the wave function belonging to these particles also guides the motion of other particles, especially when these particles are bare and the guiding way is deterministic.

*5.2 The many-worlds interpretation*

Now let's turn to the second approach to avoid wavefunction collapse, the many-worlds interpretation. Although this theory is widely acknowledged as one of the main interpretations of quantum mechanics, its many fundamental issues have not yet been solved (see Saunders et al 2010 and references therein). For example, the stuff of the many worlds, what they are made of, seems never adequately explained, nor are the worlds precisely defined. Moreover, no satisfactory role or substitute for probability has been found in the many worlds theories, and their consistency with quantum mechanics is still debatable. In the following, we will analyze whether there are many worlds according to the suggested interpretation of the wave function in terms of random discontinuous motion of particles.

In order to examine the validity of the many-worlds interpretation, it is crucial to know exactly what a quantum superposition is. No matter how to define the many worlds, they correspond to some branches of a quantum superposition after all (e.g. the branches where measuring devices obtain definite results, and in particular, observers have definite conscious experience). According to the picture of random discontinuous motion of particles, a quantum superposition exists in the form of random and discontinuous time division. For a superposition of two positions A and B of a quantum system, the system randomly and discontinuously moves between these two positions. At some random and discontinuous instants the system is in position A, and at the other instants it is in position B[8]. As a result, each position branch exists in a time sub-flow, and the sum of all these time sub-flows constitute the whole continuous time flow. In this picture of quantum superposition, it is obvious that there is only one system all along in the continuous time flow, which randomly and discontinuously moves throughout all branches of the superposition, no matter the system is a microscopic particle or a measuring device or an observer. In other words, there is only one world which instantaneous state is constantly changing in a random and discontinuous way.

This conclusion is also supported by a comparison between discontinuous motion and continuous motion. For a quantum particle undergoing discontinuous motion, the position of the particle changes discontinuously. For a classical particle, its position changes continuously. There is no essential difference between these two kinds of changes. For both cases the position of the particle is always definite at each instant, and the positions of the particle at different instants may be different. Moreover, the discontinuous change, like the continuous change, does not result in any branching process needed for creating the many worlds, because, among other reasons, the change happens all the while but the branching process only happens once. Therefore, if there is only one world in classical mechanics, so does in quantum mechanics according to the picture of random discontinuous motion of particles, no matter how the many worlds are defined.

---

[7] This conclusion also applies to the general hidden variables theories with added particle ontology.
[8] That the system is in a definite position A or B at every instant already implies that there is only one world at any time.

*5.3 A possible origin of wavefunction collapse*

The above analysis suggests that the de Broglie-Bohm theory and the many-worlds interpretation are not satisfactory solutions to the measurement problem according to our suggested interpretation of the wave function. If there are neither hidden variables nor many worlds that can explain the emergence of definite measurement results, then the collapse of the wave function is probably a real physical process, which is responsible for the transition from microscopic uncertainty to macroscopic certainty. Accordingly, the dynamical collapse theories may be in the right direction by admitting wavefunction collapse (Ghirardi 2008)[9].

However, the existing dynamical collapse theories are still phenomenological models, and they are also plagued by some serious problems such as energy non-conservation etc (Pearle 2007, 2009). In particular, the physical origin of the wavefunction collapse, including the origin of the randomness of the collapse process, is still unknown, though there are already some interesting conjectures (see, e.g. Diosi 1984; Penrose 1996). In the following, we will briefly analyze the possible origin of wavefunction collapse in terms of the random discontinuous motion of particles, and a detailed analysis will be given in another separate paper.

According to our suggested interpretation of the wave function, the wave function of a quantum particle can be regarded as an instantaneous intrinsic property of the particle that determines its random discontinuous motion. However, the wave function is not a complete description of the instantaneous state of the particle. The instantaneous state of the particle at a given instant also includes its random position, momentum and energy etc at the instant[10]. As a result, these random variables may also appear in the complete evolution equation of the instantaneous state, or in other words, they may also play a role in determining the instantaneous states at later instants in the equation. Since these variables are essentially random, their values at an instant will not influence their values at other instants in any direct way. Then these random variables can only manifest themselves in the law of motion by their influences on the evolution of the wave function. This forms a feedback in some sense; the wave function determines the probabilities of these variables assuming a particular value, while the random values of these variables at each instant also influence the evolution of the wave function in a stochastic way. Therefore, the evolution of the wave function will be governed by a revised Schrödinger equation in general, which includes the normal linear terms and a stochastic nonlinear term resulting from the influences of these random variables. It has been shown that a certain form of such stochastic evolution may lead to the right collapse of the wave function, which can explain the emergence of definite measurement results and the macroscopic world (Gao 2006).

To sum up, the existence of the collapse of the wave function seems natural according to our suggested interpretation of the wave function. On the one hand, the random discontinuous motion of particles may provide an appropriate random source to collapse the wave function, and thus it might be the physical origin of the wavefunction collapse. On the other hand, the collapse of the wave function just releases the randomness and discontinuity of motion, and as a result, the random discontinuous motion of particles can also manifest itself.

---

[9] As noted earlier, the ontology of these theories, such as mass density ontology and flashes ontology (Ghirardi, Grassi and Benatti 1995; Ghirardi 1997, 2008; Allori et al 2008), is inconsistent with our suggested interpretation of the wave function in terms of random discontinuous motion of particles. Especially, the existence of the effective mass and charge density of a quantum system seems to already exclude the mass density ontology.

[10] Although the probabilities of these variables assuming a particular value are determined by the wave function, the random values of these variables at every instant are new physical facts independent of the wave function.